\documentclass[aps,prd,twocolumn,superscriptaddress,nofootinbib]{revtex4}

\usepackage{graphicx}

\usepackage{amsmath}

\usepackage{amssymb}

\usepackage{pstricks}

\begin{document}

\title{Phantom-like behaviour in
a brane-world model with curvature effects}

\author{Mariam Bouhmadi-L\'{o}pez}
\email{mariam.bouhmadi@fisica.ist.utl.pt}
\affiliation{Centro Multidisciplinar de Astrof\'{\i}sica - CENTRA, Departamento de F\'{\i}sica, Instituto Superior T\'ecnico, Av. Rovisco Pais 1,
1049-001 Lisboa, Portugal}

\author{Paulo Vargas Moniz}
\email{pmoniz@ubi.pt}
\affiliation{Departamento de F\'{\i}sica,
Universidade da Beira Interior, Rua Marqu\^{e}s d'Avila e Bolama,
6201-001 Covilh\~{a}, Portugal}
\affiliation{Centro Multidisciplinar de Astrof\'{\i}sica - CENTRA, Departamento de F\'{\i}sica, Instituto Superior T\'ecnico, Av. Rovisco Pais 1,
1049-001 Lisboa, Portugal}

\begin{abstract}
Recent observational evidence seems to allow the possibility that
our universe may currently be under a dark energy effect of a
\emph{phantom} nature. A suitable {\em effective} phantom fluid
behaviour can emerge in brane cosmology; In particular, within the
normal non self-accelerating  DGP branch, {\em without} any exotic
matter and due to curvature effects from induced gravity. The
phantom-like behaviour is based in defining an effective energy
density that grows as the brane expands. This
effective description breaks down at some point in the past when the
effective energy density becomes negative and the effective equation
of state parameter blows up. In this paper we investigate if
the phantom-like regime can be enlarged  by the inclusion of a
Gauss-Bonnet (GB) term into the bulk. The motivation is that such a
GB component would model additional curvature effects on the brane
setting. More precisely, our aim is to determine if the GB term,
dominating and modifying  the early behaviour of the brane universe,
may eventually extend the regime of validity of
the phantom mimicry on the brane.
However, we show that the opposite occurs: the GB effect seems instead to induce
a breakdown of the phantom-like behaviour at an even smaller redshift.
\end{abstract}

\date{\today}

\maketitle

\section{introduction}\label{sec1}

In the context of cosmology, one of the most relevant astronomical
observations  of the last decade are those from distant type Ia
supernova implying that the universe is in a state of accelerated
expansion \cite{Riess:1998cb,Perlmutter:1998np} which has been
latter on  confirmed  by other observational probes
\cite{Tegmark:2003ud,Rapetti:2004aa,Spergel:2006hy,Percival:2007yw,Giannantonio:2008zi,Komatsu:2008hk}.
The fundamental nature of what is driving the cosmic acceleration is
unknown, although  many theoretical propositions have been put
forward \cite{Copeland:2006wr,Durrer:2007re}. Invoking a
cosmological constant to explain the late-time acceleration of the
universe turns out to be the most economical option which moreover
is in agreement with all the observational data
\cite{Komatsu:2008hk}. However, it turns out that the expected
theoretical value of the cosmological constant is about 120 orders
of magnitude larger than the measured one \cite{Durrer:2007re}.

Whatever it is the fuel inducing the late-time acceleration of our
universe, from a phenomenological point of view and in the framework
of general relativity, it can be described through a dark energy
component $(\rho_d,p_d)$ with an effective equation of state
$w_{\rm{eff}}=p_d/\rho_d$. The current value of $w_{\rm{eff}}$ is
extremely close to $-1$ \cite{Komatsu:2008hk}; i.e. a cosmological
constant equation of state,  but it can be larger than -1 like in
quintessence models or even smaller \cite{Komatsu:2008hk}.

The latter case disclosed above is of no lesser importance.
Quite on the contrary: If dark energy has a phantom nature, i.e.,
$w_{\rm{eff}}<-1$ \cite{Caldwell:1999ew}, then the scientific
community is facing a most considerable twofold challenge: ($i$)
explaining the cause of the recent speed up of our universe and
($ii$) also how to accommodate a phantom energy component in our
theoretical framework; i.e. what can cause  $w_{\rm{eff}}<-1$
without invoking a real phantom energy which is known to violate the
null energy condition and induce quantum instabilities\footnote{We are referring here to a phantom energy component described through a minimally coupled scalar field with the wrong kinetic term.}
\cite{Cline:2003gs}.

A possibility to mimic such a phantom-like behaviour is within  the
paradigm of string inspired brane-world models, where matter
(standard model particles) is confined on a 4-dimensional (4D)
hypersurface embedded in a higher dimensional space-time (the bulk)
and where gravity is the only interaction experiencing the full bulk
\cite{reviewRS,reviewDGP}. More precisely, Sahni and Shtanov
proposed the $\Lambda$DGP model with a phantom-like behaviour on the
brane \cite{Sahni:2002dx,Lue:2004za}, {\em without} the need of  any
matter that violates the null energy condition on the brane.

The  $\Lambda$DGP scenario is a 5D brane-world model with infrared
modifications to general relativity caused by an induced gravity
term on the brane
\cite{Dvali:2000hr,Deffayet:2000uy,IGbrane,IGearlytime,Chimento:2006ac,Lazkoz:2006gp,Lazkoz:2007zk,BouhmadiLopez:2007ts}.
The model is based on the normal or non self-accelerating branch of
the Dvali-Gabadadze-Porrati proposal (DGP)
\cite{Dvali:2000hr,Deffayet:2000uy}, which, unlike the
self-accelerating DGP branch
 is free from the ghost problem \cite{Koyama:2007za}, being the brane
filled with cold dark matter (CDM) and a cosmological constant which drives
 the late-time acceleration of the brane. The phantom-like behaviour
 is a consequence of the extra-dimension which screens the
  brane cosmological constant  and it is based in mapping the brane evolution  to that of an equivalent
4D general relativistic phantom energy model
\cite{Sahni:2002dx,Lue:2004za}.
  More precisely, the basis of this mimicry is an effective
  energy density (in the 4D general relativistic picture) corresponding to the cosmological constant
  corrected by the curvature effect due to the induced gravity
  term on the brane. This effective energy density grows as  the
brane expands and therefore effectively it behaves as a phantom
fluid; i.e. $w_{\rm{eff}}<-1$, where $w_{\rm{eff}}$ corresponds
to the ratio between the effective energy density and the effective
 pressure. The  $\Lambda$DGP model in \cite{Sahni:2002dx,Lue:2004za}
 is by far the simplest way to
 mimic a phantom-like behaviour in a brane-world setup\footnote{It is also possible to mimic a phantom-like behaviour in a consistent way in scalar-tensor dark energy models \cite{Boisseau:2000pr}.}.
 Other brane proposals aiming to produce such a mimicry
 are based on a bulk filled with matter and/or on an energy
 exchange between the brane and the bulk, therefore modifying
 the effective equation of state of dark energy on the
 brane \cite{BouhmadiLopez:2005gk}.

The effective description of the phantom  behaviour in the $\Lambda$DGP model  breaks down at
a finite redshift\footnote{We would like to stress that the
breakdown of the effective phantom description in the $\Lambda$DGP model does not
imply any sort of singularity on the brane nor in the bulk.} (cf. figure 6 and section IV for a more
detailed description); i.e. the effective energy density
 vanishes and becomes negative over a certain
redshift. When the effective energy density vanishes, the effective equation of state blows up. Given that
the phantom-like behaviour results from ($i$) induced gravity effects on
the brane causing curvature corrections and ($ii$) describing the brane model as a 4D relativistic phantom energy setup, could  the break down of the phantom-like behaviour be eliminated by considering
further curvature effects on the brane-world scenario? This is the
main question we address in this paper.

We will model such additional and new curvature effects through a
Gauss-Bonnet term (GB) in the bulk \cite{GB,GB1}. The reason behind
including this specific curvature terms is that it induces  an
ultraviolet correction  on the brane \cite{GB,GB1}, as expected from
high-energy stringy features, and therefore may modify the
phantom-like behaviour at earlier times. Eventually, affecting its
long term dynamics and even possibly preventing the mentioned break
down to occur at all. The other reason for considering such a
curvature term was anticipated recently \cite{Kofinas:2003rz,richard}: even
though the DGP model is characterised by an interesting infrared
effect of gravity occurring with respect to general relativity,
which for the self-accelerating branch can lead to a late-time
acceleration on the brane even in the absence of any exotic matter
invoked to produce the dark energy effect \cite{Deffayet:2000uy}, it
would be expected that a consistent DGP brane model would have also
ultraviolet modifications as well, associated to high-energy stringy
effects at earlier times.

This paper is therefore outlined as follows. In section \ref{sect2},
we define a brane-world model, henceforth designated as
$\Lambda$DGP-GB model, and constrain the set of parameters that
defined it in such away that the brane is currently accelerating. We
also comment on the non super-acceleration of the brane. In section
\ref{sect4} we solve the cubic Friedmann equation for the normal DGP
branch with a GB term in the bulk: The reason is to obtain an
accurate description of the effective energy density,  that will
behave like a phantom component on the brane,  which depends
explicitly on the Hubble rate. In section \ref{sect3}, we
subsequently show how a mimicry of a phantom  behaviour takes place
on the brane {\em without} considering any matter
 that violates the null energy condition on the brane.
Then, we  compare the  behaviour of the $\Lambda$DGP-GB setting
 with the behaviour found on the $\Lambda$DGP model \cite{Sahni:2002dx,Lue:2004za}.
Finally, in section \ref{sect5} we summarise and conclude. We also
present some results related to the solutions of the cubic Friedmann
equation (\ref{Friedmannnb}) in the appendix A. On the other
hand, in the appendix B, we show under which conditions the
$\Lambda$DGP model is recovered from the model we propose.


\section{Accelerating $\Lambda$DGP-GB  model
and Parameter Constraints}\label{sect2}

The generalised Friedmann equation of a
brane with induced gravity embedded in a 5D Minkowski bulk with a GB term reads\footnote{We restrict to the Friedmann equation  that has an induced gravity limit and therefore contains the DGP model \cite{richard}.} \cite{Kofinas:2003rz,richard}
\begin{eqnarray}
\left(1+\frac83 \alpha H^2\right)^2H^2=\left(r_cH^2-\frac{\kappa_5^2}{6}\rho\right)^2,
\label{Friedmann1}\end{eqnarray}
where a mirror symmetry has been assumed across the brane. In the previous equation $r_c$ is the crossover scale in the DGP model \cite{Dvali:2000hr} and has length unit. This parameter measures the strength of the induced gravity effect on the brane and is related to the 4D and 5D gravitational constants by
\begin{equation}
r_c=\frac{\kappa_5^2}{2\kappa_4^2}.
\end{equation}
On the other hand, the parameter $\alpha$
measures the strength of the GB curvature effect
on the brane and has length square unit and is positive \cite{GB}.

If $\alpha=0$, then the induced gravity in the
DGP setup modifies the late-time evolution of the brane with
respect to the standard 4D relativistic case \cite{Deffayet:2000uy}
(for an alternative approach where the induced gravity effect
corresponds to a correction to RS model at high energies see, ~e.g.,
 \cite{IGearlytime}). However, if
it is instead $r_c=0$ then  the GB term modifies the early-time evolution of the brane \cite{GB}.

Equation (\ref{Friedmann1}) can be conveniently rewritten as
\begin{eqnarray}
 H^2=\frac{\kappa_4^2}{3}\rho\pm\frac{1}{r_c}\left(1+\frac83 \alpha H^2\right)H,
\label{Friedmann2}
\end{eqnarray}
which generalises the Friedmann equation of the
self-accelerating DGP solution
\cite{Dvali:2000hr,Deffayet:2000uy} ($+$ sign in
Eq.~(\ref{Friedmann2}) with $\alpha=0$); i.e we
recover the self-accelerating solution when $\alpha=0$.
On the other hand, Eq.~(\ref{Friedmann2})   also has as a
particular solution  the DGP normal branch or non-self-accelerating solution
\cite{Dvali:2000hr,Deffayet:2000uy} ($-$ sign
in  Eq.~(\ref{Friedmann2}) when $\alpha=0$).

From now on, we restrict to the normal branch; i.e.
$-$ sign  in  Eq.~(\ref{Friedmann2}).
In addition,  we consider that the energy density
 of the brane $\rho$ corresponds to a CDM component with energy density
 $\rho_m$ and a cosmological constant $\Lambda$
\begin{equation}
\rho=\rho_m+\Lambda,
\end{equation}
with the latter driving the late-time acceleration of the brane. We will refer to this scenario as the $\Lambda$DGP-GB  model. The total energy density of the brane
is conserved and therefore also the sector corresponding to the CDM,
 which scales in the standard way with the redshift
\begin{equation}
\rho_m=\rho_{m_0}(1+z)^3.
\end{equation}
From now on a subscript $0$ stands for the observed current value of a given quantity.
Finally, the Friedmann equation on the brane can be presented as
\begin{equation}
E^2(z)=\Omega_m(1+z)^3+\Omega_\Lambda - 2\sqrt{\Omega_{r_c}}\left [1+\Omega_\alpha E^2(z)\right ]E(z),
\label{Friedmannz}
\end{equation}
where $E(z)=H/H_0$ and
\begin{equation}
\Omega_m=\frac{\kappa_4^2 \rho_{m_0}}{3H_0^2},\,\,\,\, \Omega_\Lambda=\frac{\kappa_4^2 \Lambda}{3H_0^2},\,\,\,\, \Omega_{r_c}=\frac{1}{4r_c^2H_0^2},
\end{equation}
are the usual convenient dimensionless parameters while the new parameter $\Omega_{\alpha}$ is defined as
\begin{equation}
\Omega_{\alpha}=\frac{8}{3}\alpha H_0^2.
\end{equation}

Evaluating the Friedmann equation (\ref{Friedmannz}) at $z=0$
gives a constraint on the cosmological parameters of the model
\begin{equation}
\Omega_m+\Omega_\Lambda
=1+2\sqrt{\Omega_{r_c}}\left(1+\Omega_{\alpha}\right).
\label{cosmoconstraint}
\end{equation}
For $\Omega_{\alpha}=0$ we recover the constraint
 in the $\Lambda$DGP model. The constraint (\ref{cosmoconstraint})
implies that the region $\Omega_m+\Omega_\Lambda<1$ is unphysical.
Moreover, although the brane is spatially  flat,
the previous constraint can be interpreted
in the sense
that our model constitutes a mimic of a
 closed FLRW universe in the ($\Omega_m$,\,$\Omega_\Lambda$) plane.
In particular, this is likewise to what
happens in $\Lambda$DGP, QDGP and CDGP models
\cite{Sahni:2002dx,Lue:2004za,Chimento:2006ac,Lazkoz:2006gp,BouhmadiLopez:2007ts}.
We recall  that the QDGP and CDGP models correspond
to variants of the $\Lambda$DGP scenario, where
the late-time evolution of the universe
is driven by a quiessence \cite{Chimento:2006ac}
and a Chaplygin gas \cite{BouhmadiLopez:2007ts},
respectively, instead of a  cosmological constant.
Their dark energy effect is  more dynamical
and the phantom divide (or the $w=-1$ line) crossing is possible
in the QDGP and CDGP unlike in the $\Lambda$DGP model
\cite{Chimento:2006ac,BouhmadiLopez:2007ts}.
We remind that the interest on modelling a mimicry of a phantom
divide crossing is based on the possibility (backed by recent
observational data)
that
the equation of state may have
crossed the cosmological constant barrier ($w=-1$).

Coming back to our model, if the dimensionless crossover
energy density $\Omega_{r_c}$ is the same in a $\Lambda$DGP model
and in our model (which is not necessarily the case),    then the similarities
with  a spatially closed universe are made more significant from  the GB
effect, since  $\Omega_{\alpha}>0$. Notice that this statement also applies
to the variants of the $\Lambda$DGP brane mentioned previously and their
generalisations  by the GB effect, if
the acceleration of both branes is driven by the
same sort of dark energy. In fact, in this case the appropriately modified constraint (\ref{cosmoconstraint}) would read
\begin{equation}
\Omega_m+\Omega_{\rm{DE}}
=1+2\sqrt{\Omega_{r_c}}\left(1+\Omega_{\alpha}\right),
\label{cosmoconstraint2}
\end{equation}
where $\Omega_{\rm{DE}}$ correspond to the current dimensionless energy density of dark energy on the brane which can be for example modelled by a quiessence or a Chaplygin gas.

\begin{figure}
\includegraphics[width=0.7\columnwidth]{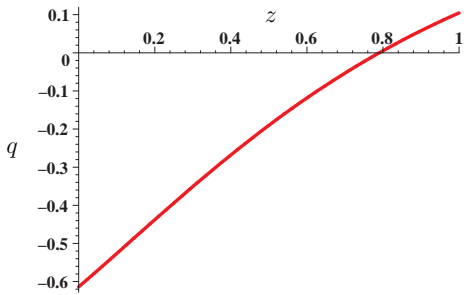}
\caption{Plot of  the deceleration parameter, \mbox{$q=-(\dot H/H^2 +1)$}, versus the redshift. The set $(\Omega_m,\Omega_\Lambda,\Omega_{r_c},\Omega_{\alpha})=(0.26,0.7602,10^{-4}, 0.01)$. As can be seen the brane accelerates at late-time when $q$ gets negative.}
\label{pq}
\end{figure}

Furthermore, by imposing that the universe is currently
accelerating; i.e. the deceleration parameter $q=-(\dot H/H^2 +1)$
is currently negative, where
\begin{equation}
q_0=-\left [1-\frac{3\Omega_m}{2+2\sqrt{\Omega_{r_c}}(1+\Omega_\alpha)}\right],
\end{equation}
we obtain another constraint
 on the set of cosmological parameters $\Omega_m$, $\Omega_{r_c}$ and $\Omega_\alpha$,
 which reads
\begin{equation}\label{constraintacce}
3\Omega_m<2+2(1+\Omega_\alpha)\sqrt{\Omega_{r_c}}.
\end{equation}
An example of the cosmological evolution of the deceleration parameter is given in Fig.~\ref{pq} where it can be seen that
the brane accelerates at late-time.

On the other hand, the modified Raychaudhuri equation follows easily from the Friedmann equation of the brane and
the conservation of the brane energy density. It  can be written as
\begin{equation}
\frac{\dot H} {H_0^2}=-\frac32\frac{\Omega_m (1+z)^3 E(z)}{E(z)+\sqrt{\Omega_{r_c}}(1+3\Omega_\alpha E^2(z))},
\label{Raychauduri}
\end{equation}
where a dot stands for the derivative respect
 to the cosmic time. The key point of  the previous
equation is that the brane never super-accelerates; i.e.
the Hubble rate decreases as the brane expands. Nevertheless,
as we will  show in section \ref{sect3}, a phantom-like behaviour takes place at
the brane: This occurs
{\em without} including any matter that violates the null energy condition.
 The phantom-like behaviour is based in defining an effective energy density
  which corresponds to a balance between the cosmological constant and
  geometrical effects encoded on the Hubble rate evolution.
  Therefore, in order to get the evolution of the effective energy
  density with the redshift it is necessary to solve the cubic
  Friedmann equation of the normal branch (Eq.~(\ref{Friedmann2})
   with (-) sign). The solutions and an  analysis  of the mentioned Friedmann equation is presented on the next section.

\section{Normal DGP branch with Gauss-Bonnet effect}\label{sect4}

Let us herein solve analytically the  Friedmann equation (\ref{Friedmann2}) for the
(-) sign for the reasons mentioned previously\footnote{The Friedmann equation (\ref{Friedmann1}) has been previously analysed in \cite{Kofinas:2003rz}. Our new analytical analysis is exclusively based on the conveniently modified cubic Friedmann equation (\ref{Friedmann2})
which make the study much easier because the discriminant of the cubic equation, $\mathcal{N}$ defined in Eq.~(\ref{NRR1}), is much simpler.}.  Concerning this aim,
it is convenient to introduce the dimensionless variables
\begin{eqnarray}
\bar H&=&\frac83 \frac{\alpha}{r_c}H = 2\Omega_\alpha\sqrt{\Omega_{r_c}}E(z), \\
\bar\rho&=&\frac{32}{27}\frac{\kappa_5^2\alpha^2}{r_c^3}\rho=4\Omega_{r_c}\Omega_\alpha^2\left[\Omega_\Lambda+\Omega_m(1+z)^3\right],\label{dimensionlessrho}\\
 b&=&\frac83\frac{\alpha}{r_c^2}=4\Omega_\alpha\Omega_{r_c}, \label{defb}
\label{dimensionlessq}
\end{eqnarray}
Then the Friedmann equation can be rewritten as
\begin{equation}
{\bar H}^3+{\bar H}^2+b\bar H-\bar\rho=0.
\label{Friedmannnb}
\end{equation}
The number of real roots is determined by the sign of the discriminant function
${\mathcal{N}}$
defined as \cite{Abramowitz}
\begin{equation}
{\mathcal{N}}= Q^3+R^2,
\label{NRR1}
\end{equation}
where $Q$ and $R$ read
\begin{equation}
Q=\frac13\left(b-\frac13\right),\quad R=\frac16 b+\frac12 \bar\rho -\frac{1}{27}.
\label{QR}
\end{equation}
It is
helpful to rewrite ${\mathcal{N}}$  as
\begin{equation}
{\mathcal{N}} =\frac14(\bar\rho-\bar\rho_1)(\bar\rho-\bar\rho_2),
\label{NRR2}
\end{equation}
where
\begin{eqnarray}
\bar\rho_1&=&-\frac13\left\{b-\frac29\left[1+\sqrt{(1-3b)^3}\right]\right\},\label{defbarrho1} \\
\bar\rho_2&=&-\frac13\left\{b-\frac29\left[1-\sqrt{(1-3b)^3}\right]\right\},
\label{defbarrho2}
\end{eqnarray}
for the analysis of the number of physical
 solutions of the modified Friedmann equation (\ref{Friedmannnb}).
If ${\mathcal{N}}$  is positive then there is a unique real solution.
On the other hand, if ${\mathcal{N}}$  is negative there are 3 real solutions. Finally,
if ${\mathcal{N}}$  vanishes, all roots are real and at least two are equal.

\subsection{Case 1: $0<b<\frac14$}

This is by far the most interesting physical case as
we expect $b$ to be small  because it is proportional to
$\Omega_{r_c}$ [see Eq.~(\ref{defb})] and the equivalent
quantity in the $\Lambda$DGP scenario is relatively small
 \cite{Lazkoz:2006gp,Lazkoz:2007zk}; We do
 not expect that $\Omega_{r_c}$ in our model to be very
 different from that  in the $\Lambda$DGP model.  Furthermore,
 from the mimicry of our model regarding a closed FLRW universe [see Eq.~(\ref{cosmoconstraint})] and the constraint on the curvature of the universe; for example from the recent WMAP 5 years data in combination with the baryon acoustic oscillations \cite{Komatsu:2008hk}, $\Omega_{r_c}$ and $\Omega_\alpha$ should be small and therefore $b$ is also expected to be small.

The analysis of this case is slightly involved.
The reason is essentially that if $0<b<\frac14$
then $\bar\rho_1$ and $\bar\rho_2$ are real
[see Eqs~(\ref{defbarrho1}) and (\ref{defbarrho2})]
 and therefore it  is not as straightforward
as in the next cases to know the number of real
solutions of the cubic Friedmann equation on $\bar H$; i.e.
to know the sign of $\mathcal{N}$ [see Eq.~(\ref{NRR2})].

More precisely, in this case $\bar\rho_2<0$ and $0<\bar\rho_1$
[see Eqs~(\ref{defbarrho1})-(\ref{defbarrho2})].
Then, the number of real roots of the cubic
Friedmann equation (\ref{Friedmannnb})
depends crucially on the minimum energy density of the brane:
\begin{equation}
\bar\rho_{\rm{min}}=4\Omega_{r_c}\Omega_\alpha^2\Omega_\Lambda;
\end{equation}
i.e. the asymptotic value of the total energy density at $z=-1$. We enumerate next the possible different situations [see Fig.~\ref{p8}]:

\begin{figure}[h]
\begin{center}
\includegraphics[width=0.7\columnwidth]{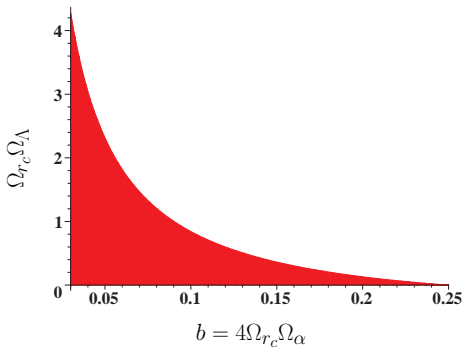}
\end{center}
\caption{The coloured area corresponds to the set ($\Omega_{r_c}$, $\Omega_\alpha$,$\Omega_\Lambda$) that does not fulfil the inequality (\ref{inequality}). This is the most likely situation as $\Omega_{r_c}$  and  $\Omega_\alpha$ are expected to be small. On the other hand, the uncoloured area correspond to the set ($\Omega_{r_c}$,$\Omega_\alpha$,$\Omega_\Lambda$) that  fulfil the inequality (\ref{inequality}).}
\label{p3}
\end{figure}

\subsubsection{$\bar\rho_1<\bar\rho_{\rm{min}}$}

The minimum energy density of the brane
is such that $\bar\rho_1<\bar\rho_{\rm{min}}$. Then
the function $\mathcal{N}$ is positive and there is a unique solution.
The condition $\bar\rho_1<\bar\rho_{\rm{min}}$ implies
\begin{equation}
-\frac{1}{3b}\left\{b-\frac29\left[1+\sqrt{(1-3b)^3}\right]\right\}<\,\,\Omega_\alpha\Omega_\Lambda,
\label{inequality}
\end{equation}
and therefore constrains the set of allowed values of $\Omega_{r_c}$, $\Omega_\alpha$ and  $\Omega_\Lambda$. In Fig.~\ref{p3} we show the set of those parameters that fulfil the inequality (\ref{inequality}) as the uncoloured area while the red coloured area corresponds to the set ($\Omega_{r_c}$, $\Omega_\alpha$, $\Omega_\Lambda$) that does not fulfil the condition (\ref{inequality}).

Finally, the expansion of the brane is described by Eq.~(\ref{H1}) or
\begin{equation}\label{barHb14unique}
\bar H_1=\frac13\left[2\sqrt{1-3b}\cosh\left(\frac{\eta}{3}\right)-1\right], 
\end{equation}
where $\eta$ is defined as
\begin{equation}
\cosh(\eta)=\frac{R}{\sqrt{-Q^3}},\quad \sinh(\eta)=\sqrt{\frac{Q^3+R^2}{-Q^3}},\label{defeta2}
\end{equation}
and $\eta_{\rm{min}}\leq \eta$. The parameter $\eta_{\rm{min}}$ is defined as in Eq.~(\ref{defeta2}) with $\bar\rho=\bar\rho_{\rm{min}}$ and this value of $\eta$ is reached at $z=-1$. It turns out that the expanding brane solution is asymptotically de Sitter in the future.
On the other hand,  at early time (large $\eta$) matter on the brane is dominated by dust though its cosmological evolution does not correspond to the standard relativistic dust case because at high redshift $\bar H \sim \bar\rho_m^\frac13$, where $\bar\rho_m$ is defined as in Eq.~(\ref{dimensionlessrho}). This is a consequence of the dominance of GB  effects at high energy. This feature applies also to the high energy regime described in the next subsection.

\subsubsection{$\bar\rho_{\rm{min}}\leq\bar\rho_1$}

The minimum energy density of the brane is such that $\bar\rho_{\rm{min}}\leq\bar\rho_1$. Consequently, the
inequality (\ref{inequality}) is not satisfied and this again restricts the set ($\Omega_{r_c}$, $\Omega_\alpha$, $\Omega_\Lambda$)
which in this case  corresponds to the coloured area in Fig.~\ref{p3}. As this figure highlights
this is the most likely situation as we expect $\Omega_{r_c}$  and  $\Omega_\alpha$ to be small for the reasons stated before.

As the energy density blue-shifts backward in times; i.e. the energy density grows backward in times we can distinguish three regimes:

\begin{itemize}

\item High energy regime: $\bar\rho_1<\bar\rho$.

\item Limiting regime: $\bar\rho=\bar\rho_1$.

\item Low energy regime: $\bar\rho_{\rm{min}}\leq\bar\rho<\bar\rho_1$.

\end{itemize}

During the high energy regime, the energy density of the brane, $\bar\rho$,  is bounded from below by $\bar\rho_1$ and therefore
the function $\mathcal{N}$ is positive or equivalently there is a unique solution of the cubic Friedmann equation (\ref{Friedmannnb}). During this regime, the expansion of the brane is described by Eq.~(\ref{barHb14unique}) where $0<\eta$ and  defined in Eq.~(\ref{defeta2}). When $\eta\rightarrow0$, the energy density of the brane approaches $\bar\rho_1$.

During the limiting regime, $\bar\rho=\bar\rho_1$. Consequently $\mathcal{N}$ vanishes and there are two solutions:
\begin{eqnarray}
\bar{H}_1&=&\,\,\,\,\frac13\left(2\sqrt{1-3b}-1\right), \label{h1}\\
\bar{H}_2&=&-\frac13\left(\sqrt{1-3b}+1\right). \label{h2}
\end{eqnarray}
The high energy regime connects with the limiting regime through $\bar{H}_1$. The negative solution $\bar{H}_2$ is not relevant physically.

\begin{figure}[h]
\begin{center}
\includegraphics[width=0.7\columnwidth]{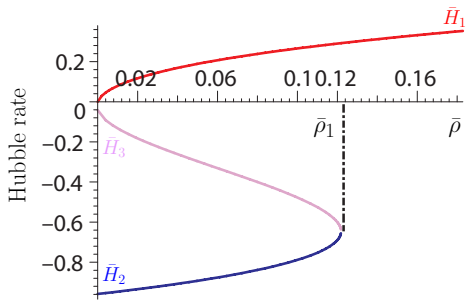}
\end{center}
\caption{Plot of the dimensionless Hubble rates $\bar H_1$, $\bar H_2$ and $\bar H_3$ against the dimensionless energy density $\bar\rho$. The red curve corresponds to $\bar H_1$. The blue curve corresponds to $\bar H_2$ and the other one to $\bar H_3$.  As can be seen it is only the solution corresponding to $\bar H_1$ that can exist in the ``far'' past and it is also the only expanding solution. If the minimum energy density of the brane $\bar\rho_{\rm{min}}$  is larger than $\bar\rho_1$ only the solution $\bar{H}_1$ exits. In the opposite case; i.e. $\bar\rho_{\rm{min}}\leq\bar\rho_1$, the three solutions exist.}
\label{p8}
\end{figure}

Finally, at the low energy regime the total energy density of the brane is bounded from above by $\bar\rho_1$. Then $\mathcal{N}$ is negative and there are 3 different solutions [see Fig.~\ref{p8}]. One of this solution corresponds to an expanding brane while the other two
corresponds to contracting branes:

It can be shown that the expanding solution ($\bar H >0$) is described by Eq.~(\ref{H1}) and more appropriately rewritten as
\begin{equation}
\bar H_1=\frac13\left[2\sqrt{1-3b}\cos\left(\frac{\theta}{3}\right)-1\right],\,\, 0 <\theta\leq\theta_{\rm{max}}
\label{solrel1}
\end{equation}
where
\begin{equation}
\cos(\theta)=\frac{R}{\sqrt{-Q^3}},\quad \sin(\theta)=\sqrt{1+\frac{R^2}{Q^3}},\label{deftheta}.
\end{equation}
For $\theta\rightarrow 0$, the energy density $\bar\rho$ approaches $\bar\rho_1$; i.e. the low energy regime is connected with the high energy regime through the solution (\ref{h1}).
On the other hand, $\theta_{\rm{max}}$ is defined as in Eq.~(\ref{deftheta})
with $\bar\rho=\bar\rho_{\rm{min}}$ where the brane reaches its asymptotic de Sitter regime
at $z=-1$. Notice that as matter redshifts on the brane, the angle $\theta$ gets larger. On the other hand, in this model, if the cosmological constant vanishes then the maximum angle $\theta$ is given by $\theta_0$ where
\begin{equation}
\cos\left(\frac{\theta_0}{3}\right)=\frac{1}{2\sqrt{1-3b}},
\label{deftheta0}
\end{equation}
and therefore, the Hubble rate vanishes. This feature signals that this brane solution does not corresponds to a self-accelerating brane.

For completeness, we write down the remaining two solutions of the
Friedmann equation (\ref{Friedmannnb}) when $0<b<1/4$ and $\bar\rho<\bar\rho_1$.
As it was anticipated before,  these solutions describe contracting branes and correspond to the solutions given in Eqs.~(\ref{H2}) and  (\ref{H3}). They read
\begin{eqnarray}
\bar{H}_2&=&-\frac13\left[2\sqrt{1-3b}\cos\left(\frac{\pi-\theta}{3}\right)+1\right],\, 0 <\theta\leq\theta_{\rm{max}},\nonumber \\
\\
\bar{H}_3&=&-\frac13\left[2\sqrt{1-3b}\cos\left(\frac{\pi+\theta}{3}\right)+1\right],\, 0 <\theta\leq\theta_{\rm{max}},\nonumber \\ \label{solrel2}
\end{eqnarray}
respectively. Unlike the solution $\bar H_1$,
these two solutions are contracting because $\bar{H}_2$
 and $\bar{H}_3$ are negative. It can be shown that $\bar{H}_2\leq\bar{H}_3$.
Both solutions approaches the same Hubble rate in the past at $\theta=0$ corresponding to the limiting solution (\ref{h2}).
Finally, if there is no cosmological constant on the brane,
then in the far future (at $z=-1$)  the angle $\theta$ is given by Eq.~(\ref{deftheta0}), where  $\bar{H}_2$ and $\bar{H}_3$ approach constant negative values.

Before ending we would like to point out that it is only the
expanding branch with Hubble rate $\bar H_1$ that has a phantom-like
behaviour which we will describe in the next section.  An example of
the three different solutions of the Friedmann equation
(\ref{Friedmannnb}) for $0<b<\frac14$ and
$\bar\rho_{\rm{min}}\leq\bar\rho_1$ can be seen in Fig.~(\ref{p8}) .

\subsection{Case 2: $\frac14\leq b<\frac13$ }

If\footnote{The
parameter $\bar\rho_1$ vanishes at $b=1/4$.}
$\frac14\leq b<\frac13$ then $\bar\rho_1\leq0$ and $\bar\rho_2<0$  which
implies that $\mathcal{N}$ is positive because the total energy
density of the brane $\bar\rho$ is positive.
Consequently, there is a unique real solution
and the cosmological evolution of the brane is unique. The dimensionless Hubble parameter is
given in Eqs.~(\ref{barHb14unique})-(\ref{defeta2}) and $\eta_{\rm{min}}$ is defined as in Eq.~(\ref{defeta2}) with $\bar\rho=\bar\rho_{\rm{min}}$.
When $\eta$ approaches its minimum value,  the Hubble rate is constant and positive;
i.e. the brane is asymptotically de Sitter and its expansion is dominated by the cosmological constant.
If all matter on the brane redshifts and the total energy
density on the brane vanishes at $z=-1$ then at $\eta=\eta_0$, where
\begin{equation}
\cosh\left(\frac{\eta_{\rm{0}}}{3}\right)=\frac{1}{2\sqrt{1-3b}},
\end{equation}
the Hubble rate vanishes. This feature is in agreement with the fact that this solution does not
correspond to a self-accelerating branch.

At high energy (large $\eta$) matter on the brane is dominated by dust. However, its cosmological evolution does not correspond to the standard
relativistic dust case because at high redshift $\bar H \sim \bar\rho_m^\frac13$ due to the GB effects at high energy.

\subsection{Case 3: $b=\frac13$}

This constitutes a  marginal case where $b=1/3$; i.e.
$\Omega_{\alpha}=1/(12\Omega_{r_c})$.
The modified Friedmann  Eq.~(\ref{Friedmannnb}) has a
unique real solution because  $\mathcal{N}>0$. This can
be noticed easily by realising that $\bar\rho_1=\bar\rho_2 <0 $   when
$b=1/3$ and therefore the right hand side of Eq.~(\ref{NRR2}) is always positive.
The dimensionless Hubble parameter given in
Eq.~(\ref{H1}) can be rewritten in a simple way as
\begin{equation}
\bar H=\frac13\left[\left(1+27\bar\rho\right)^{\frac13}-1\right].
\end{equation}
The brane is therefore
asymptotically de Sitter in the future ($\bar\rho\rightarrow\bar\rho_{\rm{min}}$). At high energy/earlier time,
the matter on the brane is dominated by dust although
 the dimensionless Hubble parameter redshift as $\bar\rho_m^\frac13$.

\subsection{Case 4: $\frac13<b$}
In this case, Eq.~(\ref{Friedmannnb}) has  a unique real solution because  ${\mathcal{N}}>0$ as $\bar\rho_1$ and $\bar\rho_2$
are complex conjugates when $\frac13<b$ and therefore the right hand side of Eq.~(\ref{NRR2}) is always positive.

The dimensionless Hubble parameter is given in Eq.~(\ref{H1}) and can be rewritten as
\begin{equation}
\bar H=\frac13\left[2\sqrt{3b-1}\sinh\left(\frac{\eta}{3}\right)-1\right], \quad \eta_{\rm{min}}\leq \eta,
\end{equation}
where now $\eta$ fulfils
\begin{equation}
\cosh(\eta)=\sqrt{1+\frac{R^2}{Q^3}},\quad \sinh(\eta)=\frac{R}{\sqrt{Q^3}},\label{defeta}
\end{equation}
and the parameter $\eta_{\rm{min}}$ is defined as
in Eq.~(\ref{defeta}) with
$\bar\rho=\bar\rho_{\rm{min}}$.
The brane is therefore asymptotically de Sitter ($z=-1$ corresponds to $\eta$
approaching $\eta_{\rm{min}}$) though there is no self-accelerating solution.

For the sake of completeness, we point out that if all matter
on the brane redshifts and the total energy density on the brane vanishes at $z=-1$ then at this redshift $\eta=\eta_0$,
where the Hubble rate vanishes, in agreement with the fact that this solution does not correspond to
a self-accelerating branch, and where $\eta_0$ satisfies
\begin{equation}
\sinh\left(\frac{\eta_{\rm{0}}}{3}\right)=\frac{1}{2\sqrt{3b-1}}.
\end{equation}

Finally, we have that at high energy (large $\eta$) the brane is dust dominated although its cosmological evolution does not correspond to the standard relativistic dust case because $\bar H \sim \bar\rho_m^\frac13$ where $\bar\rho_m$ is defined as in Eq.~(\ref{dimensionlessrho}).

The asymptotic de Sitter regime of the brane in all the cases numerated is due to
the presence of a cosmological constant on the 4D
hypersurface (unlike the self-accelerating solutions \cite{Kofinas:2003rz}).
In addition, the fact that the brane is asymptotically de Sitter
implies that there is no big rip singularity in the future despite
that the brane has a phantom-like behaviour at late-time as we show
in the next section.

Before proceeding into discussing how a phantom-like behaviour
takes place on the brane framework we use, there is a point worthy
to emphasise. The  Gauss-Bonnet parameter   can, in principle, have an
arbitrary value. However, being also considered as a perturbative
term arising from string theory, it is sensible that  within the
model discussed here, realistic cosmological solutions should
coincide with the $ \Lambda DGP$ cosmology in the limit
$\alpha\rightarrow 0$. We therefore present in the appendix B how
this is the case for one of the herein found solutions (Eq.~(\ref{solrel1})) and  under
which conditions the $\Lambda$DGP model is recovered from the
model we propose. This procedure follows the analysis in Refs.
\cite{GB1,Kofinas:2003rz}. Moreover, it will be the solution
(\ref{solrel1}) that will be employed in the next section.

\section{Phantom-like behaviour on the brane and Gauss-Bonnet effect}\label{sect3}

\begin{figure}
\includegraphics[width=0.7\columnwidth]{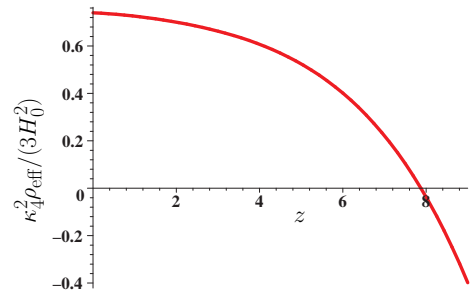}
\caption{Plot of  the dimensionless effective energy density versus the redshift. The set $(\Omega_m,\Omega_\Lambda,\Omega_{r_c},\Omega_{\alpha})=(0.26,0.7602,10^{-4}, 0.01)$ which corresponds to $b=4\times 10^{-6}$ where $b$ is defined in Eq.~(\ref{defb}). The chosen values for $(\Omega_m,\Omega_\Lambda,\Omega_{r_c})$ are in agreement with the best observational fit of the $\Lambda$DGP model \cite{Lazkoz:2007zk} and suitable for
our model as  we expect GB effects in our model to correspond to small corrections to the $\Lambda$DGP scenario.}
\label{prho}
\end{figure}

The phantom-like behaviour on the brane
 is based in defining a corresponding  effective energy density
$\rho_{\rm{eff}}$ and an effective equation of state with parameter
$w_{\rm{eff}}$. More precisely, the effective description
is inspired in writing down the modified Friedmann
equation of the brane as the  usual relativistic
Friedmann equation so that
\begin{eqnarray}
H^2=
\frac{\kappa_4^2}{3}(\rho_m+\rho_{\rm{eff}});
\label{Friedmann3}
\end{eqnarray}
i.e. to map the brane evolution in Eq.~(\ref{Friedmannz}) to the equivalent 4D general relativistic phantom cosmology with Friedmann equation (\ref{Friedmann3}). In the previous equation the effective energy density  $\rho_{\rm{eff}}$ reads
\begin{eqnarray}
\rho_{\rm{eff}}&=&\Lambda-\frac{3}{\kappa_4^2r_c}\left(1+\frac83\alpha H^2\right)H, \nonumber \\
 &=&\frac{3 H_0^2}{\kappa_4^2}\left[\Omega_\Lambda-2\sqrt{\Omega_{r_c}}(1+\Omega_\alpha E^2(z))E(z)\right].
\label{rhoeff}
\end{eqnarray}
This effective energy density corresponds to a balance between the cosmological constant and geometrical effects encoded on the Hubble parameter. On the other hand, gravity leakage at late-time screens the cosmological constant like in the $\Lambda$DGP scenario \cite{Sahni:2002dx,Lue:2004za}. This phantom-like behaviour is obtained without any matter violating the null energy condition and without any super-acceleration on the brane.
We stress that the dependence of $\rho_{\rm{eff}}$ on the redshift is known analytically by means of the different solutions $\bar{H}$ of the cubic Friedmann equation presented on the previous section.

In Fig.~\ref{prho} we show an example of the evolution of the dimensionless effective energy density $\kappa_4^2\rho_{\rm{eff}}/(3 H_0^2)$. In this example $b=4\times 10^{-6}$. Our choice for the values of $b$ is based on the fact that observationally the most favourable  set of solutions $\bar{H}_1$ are those such that \mbox{$0<b<\frac14$} and $\bar\rho_{\rm{min}}<\bar\rho_1$ (we refer the reader  to the previous section). Furthermore, the chosen value $b$ is obtained for $\Omega_{r_c}=10^{-4}$ which is in agreement with the best fit of the $\Lambda$DGP model \cite{Lazkoz:2007zk} and suitable for
our model as  we expect GB effects in our model to correspond to small corrections to the $\Lambda$DGP setup.

\begin{figure}
\includegraphics[width=0.7\columnwidth]{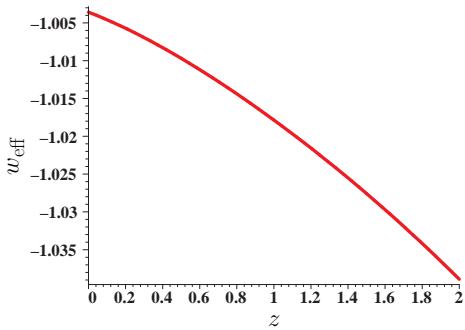}
\caption{Plot of  the effective equation of state  versus the redshift. The set $(\Omega_m,\Omega_\Lambda,\Omega_{r_c},\Omega_{\alpha})=(0.26,0.7602,10^{-4}, 0.01)$ the same considered in Fig.~\ref{prho}.}
\label{pweff}
\end{figure}

As in  the $\Lambda$DGP model,
it is possible to define an effective
equation of state or
parameter $w_{\rm{eff}}$ associated to  the effective energy density as
\begin{equation}
\dot\rho_{\rm{eff}}+3H(1+w_{\rm{eff}})\rho_{\rm{eff}}=0.
\end{equation}
This effective equation of state is defined in analogy with the standard relativistic case.
Then using Eq.~(\ref{rhoeff}), we obtain
\begin{eqnarray}
1+w_{\rm{eff}}&=&\frac{1}{\kappa_4^2r_c}\;\frac{\dot H(1+8\alpha H^2)}{H\rho_{\rm{eff}}},\label{weff} \\
&=&\frac23\frac{\dot H/H_0^2\sqrt{\Omega_{r_c}}\left(1+3\Omega_\alpha E^2(z)\right)}{E(z)\left[\Omega_\Lambda-2\sqrt{\Omega_{r_c}}\left(1+\Omega_\alpha E^2(z)\right) E(z)\right]}.\nonumber
\end{eqnarray}
Because the brane never super-accelerates,
i.e.  $\dot H <0$ [see Eq.~(\ref{Raychauduri})],
we can then conclude that $\rho_{\rm{eff}}$
mimics the behaviour of a phantom energy component on the brane:
I.e. $1+w_{\rm{eff}}<0$,
as long as the effective energy density $\rho_{\rm{eff}}$ is positive [see Figs.~\ref{prho} and \ref{pweff}].

\begin{figure}
\includegraphics[width=0.8\columnwidth]{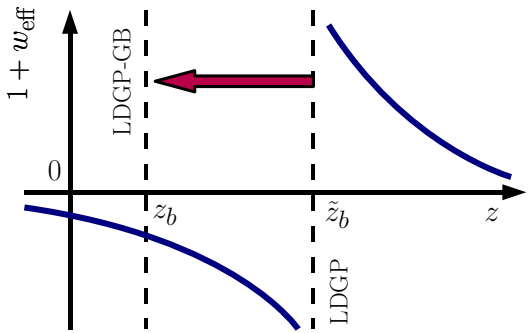}
\caption{In this plot we show how the phantom-like behaviour of a
$\Lambda$DGP model is modified by the effect of a GB curvature term
in the bulk. The effect is schematically shown by the red arrow.
Hence, the GB effect translates into a sooner breakdown of the
effective phantom-like picture; i.e. $z_b<\tilde z_b$.  The
redshifts values $\tilde z_b$ and $z_b$ correspond to the moments
when the effective energy density vanishes and becomes negative in
the pure $\Lambda$DGP and the modified one  with GB effect in the
bulk, respectively.} \label{GB-effect}
\end{figure}

We would like to point out that a phantom energy component
can be defined in two ways: ($i$) any matter such that
its equation of state fulfils $p/\rho=w<-1$ or
($ii$) any matter whose energy density grows when the universe expands.
Both definitions are equivalent as
long as the universe expands and the
energy density of the phantom component
 is positive. Here, we are assuming that
the phantom energy is defined as in ($i$) and it turns out that
the definition ($ii$) is automatically  satisfied.
In this model, the opposite situation
does not hold, in other words if condition ($ii$) is satisfied (which is always the case in this model) it does not imply that condition ($i$) is fulfilled. The reason behind it is that
$\dot\rho_{\rm{eff}}$ is always positive,
\begin{equation}\label{Hzbm}
\dot\rho_{\rm{eff}}=-\frac{3}{\kappa_4^2r_c}\dot H(1+8\alpha H^2),
\end{equation}
and therefore $\rho_{\rm{eff}}$  always grows
as the brane expands independently of its sign.
However, $1+w_{\rm{eff}}$ changes its sign, although
in an abrupt way, when $\rho_{\rm{eff}}$ vanishes
and becomes negative [see Eq.~(\ref{weff}) and Fig.~\ref{GB-effect}].

When the effective energy $\rho_{\rm{eff}}$ vanishes, the mimicry of the phantom behaviour breaks down; i.e. the mapping between the $\Lambda$DGP-GB model and the 4D relativistic phantom cosmology  model with Friedmann  Eq.~(\ref{Friedmann3}) is no longer valid , although the $\Lambda$DGP-GB model  is well defined at any redshift.
At the redshift $z_b$, where $\rho_{\rm{eff}}(z_b)=0$, the Hubble rate is constrained to fulfil [see Eq.~(\ref{rhoeff})]
\begin{equation}
2\sqrt{\Omega_{r_c}}\left[1+\Omega_\alpha E^2(z_b)\right]E(z_b)-\Omega_\Lambda=0.
\end{equation}
On the other hand, the Friedmann equation (\ref{Friedmann3}) implies
\begin{equation}\label{eq45}
E^2(z_b)=\Omega_m(1+z_b)^3.
\end{equation}
By combining the last two equations, we find that the redshift at which the mimicry of a phantom behaviour breaks down reads \cite{Abramowitz}
\begin{equation}
z_b=\left[\frac{S_+-S_-}{2\sqrt{\Omega_{r_c}\Omega_m}}\right]^{\frac23}-1,
\label{zb}
\end{equation}
where
\begin{equation}
S_{\pm}=\left[\sqrt{\left(\frac43\frac{\Omega_{r_c}}{\Omega_\alpha}\right)^3+\left(2\frac{\Omega_{r_c}}{\Omega_\alpha}\Omega_\Lambda\right)^2}\pm2\frac{\Omega_{r_c}}{\Omega_\alpha}\Omega_\Lambda\right]^{\frac13}.
\end{equation}

When the dimensionless energy density, $\Omega_{\alpha}$, associated to the GB geometrical effects is much smaller than
$\Omega_{r_c}$, the redshift $z_b$ fulfils
\begin{equation}
2\sqrt{\Omega_m\Omega_{r_c}}(1+z_b)^{\frac32}=\Omega_{\Lambda}-\frac14\Omega_{\Lambda}^3\frac{\Omega_\alpha}{\Omega_{r_c}}+O^2\left(\frac{\Omega_\alpha}{\Omega_{r_c}}\right).
\end{equation}
Therefore, in the limiting situation $\Omega_{\alpha}=0$, we recover that the phantom-like description breaks down also in the $\Lambda$DGP setup at same point in the past\footnote{We use a
tilde to define quantities in the $\Lambda$DGP model and hence
distinguish them from the ones used in our model.}:
\begin{equation}
2\sqrt{\Omega_{\tilde m}\Omega_{\tilde r_c}}(1+\tilde{z}_b)^{\frac32}=\Omega_{\tilde\Lambda}.
\label{tildezb}
\end{equation}
In fact, the effective phantom-like description
breaks down in the $\Lambda$DGP scenario when  the analogues effective energy density vanishes
and therefore the effective equation of state parameter blows up \cite{Sahni:2004fb} [see Eq.~(\ref{weff}) with $\Omega_{\alpha}=0$]. This again points to the fact the mapping between the $\Lambda$DGP model and the 4D relativistic phantom cosmology model with Friedmann expansion (\ref{Friedmann3}) (for $\alpha=0$) is no longer valid, although the $\Lambda$DGP brane description remains valid\footnote{We thank Y. Shtanov for pointing out this to us.}.

The above behaviour raises the following possibility:
Can the phantom behaviour break down in our model at a redshift $z_b$
(\ref{zb}), such that $z_b>\tilde{z}_b$?
Our   motivation is that such a GB component would model additional
curvature effects on the brane setting. The subsequent  aim is then
to determine if the GB term, even if dominating  and modifying the
early behaviour of the brane universe, may eventually extend its
effect towards later times. Thus, could
the regime of validity of the phantom mimicry in the
$\Lambda$DGP setting be extended  by considering further
curvature effects on the brane-world scenario?

\begin{figure*}
\includegraphics[width=0.9\textwidth]{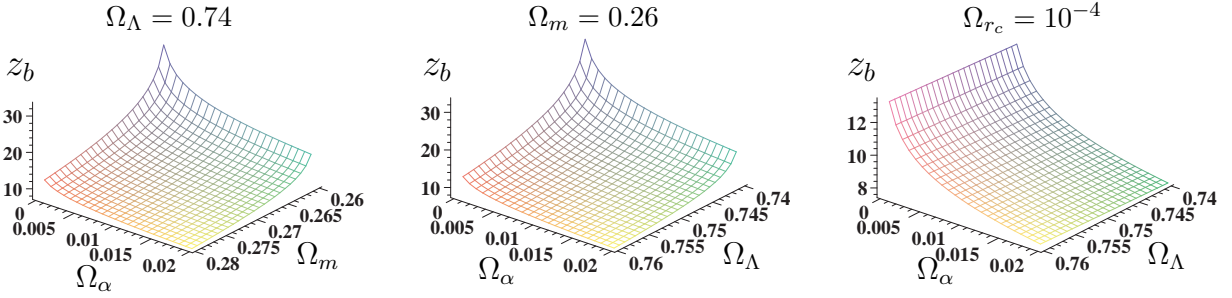}
\caption{Plot of the redshift  $z_b$ at which the phantom-like behaviour in the $\Lambda$DGP-GB model breaks down. When $\Omega_{\alpha}=0$, we obtain the redshift  $\tilde z_b$ at which the phantom-like behaviour in the $\Lambda$DGP scenario breaks down. As can be noticed
$z_b<\tilde z_b$.}
\label{pzbbreak}
\end{figure*}

The answer will depend on the cosmological parameters
that characterise both models, three
 in the $\Lambda$DGP scenario namely
($\Omega_{\tilde m}$,$\Omega_{\tilde\Lambda}$,$\Omega_{\tilde{r}_c}$)
and four in the model we are analysing namely
($\Omega_m$,$\Omega_\Lambda$,$\Omega_{r_c}$,$\Omega_\alpha$).
In order to be able to give a definite
answer, we will first assume three different cosmological situations. More general situation will be analysed by means of three dimensional plots (see Fig.~\ref{pzbbreak}).

\subsection{Fixed $\Omega_{m}$ and $\Omega_{\Lambda}$}

We first assume the same amount of dark matter $\Omega_m$ and dark energy $\Omega_{\Lambda}$ in the $\Lambda$DGP model and its twin with GB effect. Therefore, the difference between the $\Lambda$DGP scenario and $\Lambda$DGP-GB in this case is  encoded on the geometrical effects quantified through the crossover scale and the GB parameter.
By using the constraint equation (\ref{cosmoconstraint}), it turns out
that the dimensionless energy density associated  to the crossover scale cannot be the same in both models. In fact, the dimensionless energy density related to the crossover scale $\Omega_{\tilde{r}_c}$ in the $\Lambda$DGP model is related to $\Omega_{r_c}$ in the $\Lambda$DGP-GB model by
\begin{equation}
\Omega_{\tilde{r}_c}=\Omega_{r_c}(1+\Omega_{\alpha})^2.
\label{relcrossover}
\end{equation}
Therefore, in this situation the
crossover scale in the $\Lambda$DGP
model will be larger than in the  $\Lambda$DGP scenario
with a GB term in the bulk.
Let ${\tilde {z}}_b$ and ${z}_b$
be the redshift at which the effective energy
density vanishes  in the $\Lambda$DGP setup and in
our scenario, respectively. Then,  as the effective energy density vanishes at those redshifts
Eq.(\ref{rhoeff}) implies
\begin{eqnarray}
\Omega_{\Lambda}&=&2\sqrt{\Omega_{r_c}}\left [1+\Omega_\alpha E^2(z_b)\right ]E(z_b),\nonumber \\
&=&2\sqrt{\Omega_{\tilde r_c}}\tilde E(\tilde z_b),
\end{eqnarray}
which by using Eq.~(\ref{relcrossover}) translates into
\begin{equation}
\frac{\tilde E(\tilde z_b)}{E(z_b)}=\frac{1+\Omega_\alpha E^2(z_b)}{1+\Omega_\alpha}.
\end{equation}
Now we recall that $1<E(z_b)$ as the brane do not super-accelerate  and $0<z_b$.
Therefore, the right hand side of the previous equation
is larger than one, which implies that $z_b<\tilde z_b$ because of Eq.~(\ref{eq45}) and the analogous relation
\begin{equation}\label{eq53}
\tilde E(\tilde z_b)=\Omega_{\tilde{m}} (1+\tilde z_b)^3,
\end{equation}
in the $\Lambda$DGP model.

In conclusion, the phantom-like  behaviour
of the $\Lambda$DGP model with GB effect
breaks down at a smaller redshift than the phantom-like
behaviour on the $\Lambda$DGP model under the assumption that
$\Omega_m$ and $\Omega_{\Lambda}$ are the same
in the $\Lambda$DGP setup with and without GB  effect in the bulk [see Fig.~\ref{GB-effect}].

\subsection{Fixed $\Omega_{\Lambda}$ and $\Omega_{r_c}$}

We next assume the same amount of dark energy $\Omega_{\Lambda}$  and the same dimensionless energy density $\Omega_{r_c}$ in both models. Then, the constraint equation (\ref{cosmoconstraint}) implies that the amount of dark matter is slightly larger in the $\Lambda$DGP-GB brane than in the $\Lambda$DGP brane because of the GB effects
\begin{equation}\label{relom}
\Omega_m-\Omega_{\tilde{m}}=2\sqrt{\Omega_{r_c}}\Omega_\alpha.
\end{equation}
At the redshifts $\tilde z_b$ and ${z}_b$ the effective energy densities in the $\Lambda$DGP and the $\Lambda$DGP-GB models  vanish, respectively, and Eq.~(\ref{rhoeff}) implies
\begin{equation}\label{eq55}
\frac{\tilde E(\tilde z_b)}{E(z_b)}={1+\Omega_\alpha E^2(z_b)}>1,
\end{equation}
because we have assumed the same $\Omega_{\Lambda}$ and $\Omega_{r_c}$ in both models. Now, Eqs (\ref{eq45}), (\ref{eq53}) and (\ref{eq55}) imply
\begin{equation}
\left(\frac{1+\tilde{z}_b}{1+z_b}\right)^3>\frac{\Omega_m}{\Omega_{\tilde{m}}}.
\end{equation}
Therefore, $\tilde z_b > z_b$ because the amount of dark matter is larger in the $\Lambda$DGP-GB brane than in the $\Lambda$DGP brane [see Eq.~(\ref{relom})]. In conclusion, we have that the phantom-like behaviour breaks down sooner in the $\Lambda$DGP-GB model than in the $\Lambda$DGP setup [see Fig.~\ref{GB-effect}].

\subsection{Fixed $\Omega_{m}$ and $\Omega_{r_c}$}

In this case we assume that both models have the same amount of dark matter $\Omega_m$ and the same dimensionless energy density $\Omega_{r_c}$. Now, we use Eq.~(\ref{cosmoconstraint}) to constrain the remaining free parameters of both models
\begin{equation}\label{relol}
\Omega_{\Lambda}-\Omega_{\tilde\Lambda}=2\sqrt{\Omega_{r_c}}{\Omega_\alpha}.
\end{equation}
The previous constraint implies that the amount of dark energy would be slightly larger in the $\Lambda$DGP-GB model than in the $\Lambda$DGP model in this case. Because the effective energy density $\rho_{\rm{eff}}$ vanishes at
$\tilde z_b $ and $z_b$, it  implies a constraint on how different can be the amount of dark energy in both models [see Eq.~(\ref{rhoeff})]
\begin{eqnarray}\label{eq58}
\Omega_{\Lambda}-\Omega_{\tilde\Lambda}=2\sqrt{\Omega_{r_c}}\left[E(z_b)-\tilde E(\tilde z_b)\right]+2\sqrt{\Omega_{r_c}}
{\Omega_\alpha}E^3(z_b).\nonumber \\
\end{eqnarray}
By combining Eqs.~(\ref{relol}), (\ref{eq58}) and recalling that $z_b>0$, we can conclude that
\begin{equation}
\tilde E(\tilde z_b)-E(z_b)={\Omega_\alpha}\left[E^3(z_b)-1\right]>0;
\end{equation}
i.e. the dimensionless Hubble rate $\tilde E(\tilde z_b)$ is larger than the dimensionless Hubble rate $E(z_b)$. Therefore,  $z_b<\tilde z_b$ [see Eqs~(\ref{eq45}) and (\ref{eq53})]; i.e. we reach a similar conclusion to the one presented in  the previous subsections [see Fig.~\ref{GB-effect}].

So, far we have compared the redshifts $z_b$, $\tilde z_b$ where the mimicry of the phantom behaviour breaks down in the
$\Lambda$DGP-GB and $\Lambda$DGP models, respectively, under the assumption that two of the parameters that characterise both models are equals. We relax this condition in the next subsection.

\subsection{More general situations}

In order to compare further the redshifts \mbox{$\tilde z_b=\tilde z_b(\Omega_{\tilde\Lambda},\Omega_{\tilde m},\Omega_{\tilde r_c})$} and \mbox{$z_b=z_b(\Omega_\Lambda,\Omega_m,\Omega_{r_c},\Omega_\alpha)$} at which the phantom-like behaviour breaks down in the $\Lambda$DGP and  $\Lambda$DGP-GB models, respectively, we relax the conditions imposed in the previous subsection.

We consider that only one of the parameters ($\Omega_\Lambda$, $\Omega_m$, $\Omega_{r_c}$) is fixed in the $\Lambda$DGP-GB brane. We further assume that the given value of the fixed parameter is very close to  that obtained by constraining the $\Lambda$DGP model with
observational data ($H(z)$, CMB shift parameter and SNIa data)\footnote{As far as we know, the DGP-GB model has been  constrained by cosmological  observations in Ref.~\cite{He:2007zf}. However, it turns out that the mentioned paper deals with the self-accelerating branch of the model which is different from the one we explore in our paper (the one that contains the $\Lambda$DGP-GB model). The model analysed in  Ref.~\cite{He:2007zf} reduces to the self-accelerating DGP branch when $\alpha\rightarrow 0$ while our model reduces to the $\Lambda$DGP model which is contained in the normal DGP branch.  Therefore, as we are considering the GB term as a perturbation of the $\Lambda$DGP model, we take as a good approximation to consider the cosmological parameters of the $\Lambda$DGP-GB model close to those of the original  $\Lambda$DGP model.} \cite{Lazkoz:2007zk}
\begin{equation}\label{eq60}
\Omega_m=0.26\pm0.05, \quad \Omega_{r_c}\leq 0.05.
\end{equation}
Our second assumption is based on the fact that we are considering the GB effects as small corrections to the $\Lambda$DGP model. Then, once we have fixed the value of one of the parameters ($\Omega_\Lambda$, $\Omega_m$, $\Omega_{r_c}$), there remain only three free parameters in the $\Lambda$DGP-GB model. One of these parameters can be fixed by means of the  cosmological constraint equation (\ref{cosmoconstraint}). Therefore, at the end we are left with  only two free parameters and as we know analytically $z_b$ [see Eq.~(\ref{zb})] we can do a three dimensional plot of $z_b$ in terms of the two free parameters. Our  results are shown in Fig.~\ref{pzbbreak}:

On the left hand side plot in Fig.~\ref{pzbbreak}, we have chosen $\Omega_\Lambda=0.74$, its best fit value for the $\Lambda$DGP model \cite{Lazkoz:2007zk}. Then for a given set $(\Omega_{m},\Omega_{\alpha})$, $\Omega_{r_c}$ is fixed by means of the constraint (\ref{cosmoconstraint}). In this plot $\tilde z_b$ is retrieved when $\Omega_{\alpha}=0$. As can be seen for a given value $\Omega_{m}$ the largest $z_b$ corresponds to $\Omega_{\alpha}=0$; i.e. $z_b<\tilde z_b$.

On the middle plot in Fig.~\ref{pzbbreak}, we have followed a similar procedure. We have chosen $\Omega_m=0.26$, its best fit value for the $\Lambda$DGP model \cite{Lazkoz:2007zk}. Then for a given set $(\Omega_{\Lambda},\Omega_{\alpha})$, $\Omega_{r_c}$ is obtained by means of the constraint (\ref{cosmoconstraint}). In this plot $\tilde z_b$ is retrieved when $\Omega_{\alpha}=0$. As can be seen for a given value $\Omega_{\Lambda}$ the largest $z_b$ corresponds to $\Omega_{\alpha}=0$; i.e. $z_b<\tilde z_b$.

Finally, on the right hand side plot in  Fig.~\ref{pzbbreak}, we have chosen $\Omega_{r_c}=10^{-4}$ which fulfils the constraint (\ref{eq60}). For a given set $(\Omega_{\Lambda},\Omega_{\alpha})$, $\Omega_{m}$ is fixed by means of the constraint (\ref{cosmoconstraint}). In this plot $\tilde z_b$ is retrieved when $\Omega_{\alpha}=0$. It is clear form the plot that for a given value $\Omega_{\Lambda}$ the largest $z_b$ corresponds to $\Omega_{\alpha}=0$; i.e. $z_b<\tilde z_b$.

Hence, we have that in all the situations \textbf{A-D} analysed
above the phantom-like behaviour in the $\Lambda$DGP-GB brane breaks
down sooner (smaller redshift) than in the $\Lambda$DGP brane. In
more precise terms, the phantom-like behaviour in the $\Lambda$DGP
is regular  in the interval  $[0,\tilde{z}_b)$ but breaks down for
higher redshifts. On the other hand, the phantom-like behaviour in the $\Lambda$DGP-GB scenario works in
the interval $[0, z_b)$ but fails  for higher redshifts. Because
$z_b<\tilde{z}_b$, the GB effect actually  makes smaller the
interval $[0,\tilde{z}_b)$ and therefore it does not help to improve
the situation. Please note ($i$) we are  analysing the phantom-like behaviour through the mapping between
the $\Lambda$DGP/$\Lambda$DGP-GB models and the 4D relativistic phantom cosmology models with Friedmann Eq.~(\ref{Friedmann3}) from  today ($z=0$)  till $\tilde{z}_b$/$z_b$ and (ii) the break down of the phantom-like behaviour does not imply any sort of  singularity in the brane-world models.

\section{Summary and conclusion}\label{sect5}

In this paper we have analysed in some detail the $\Lambda$DGP-GB
model which corresponds to a 5D brane-world model where the bulk is
a 5D Minkowski space-time. The model contains  a GB term in the bulk
\cite{GB,GB1} and an induced gravity term on the brane
\cite{IGbrane,Kofinas:2003rz}. Our analysis was performed for the
normal or non self-accelerating branch which we have assumed to be
filled by CDM and a cosmological constant, the latter driving the
late-time acceleration of the brane. We have shown how the brane
accelerates at late-times (cf. Eq.~(\ref{constraintacce}) and
Fig.~\ref{pq}).

The attractive and promising feature of this model is the role of
the extra-dimension: It induces a mimicry of a phantom behaviour
\emph{without} resorting to any matter that violates the null energy
condition on the brane. This phantom-like behaviour happens without
any super-acceleration of the brane (see Eq.~(\ref{Raychauduri}))
and, therefore, the brane does not hit a big rip singularity in its
future. Indeed, the brane is asymptotically de Sitter. This regime
is reached when the cosmological constant dominates over the CDM
component. Our model reduces to the $\Lambda$DGP scenario
\cite{Sahni:2002dx,Lue:2004za} if the GB corrections in the bulk are
put aside.

Our motivation for considering a GB correction to the $\Lambda$DGP
is threefold: ($i$) it is known that the mimicry of a phantom
behaviour on the $\Lambda$DGP model breaks down at some point in the
past \cite{Sahni:2004fb}. This happens when the mapping between the $\Lambda$DGP and
the 4D relativistic phantom cosmology model breaks down. More precisely,
 when the effective energy density that mimics the phantom-like behaviour gets negative  and
therefore the corresponding effective equation of state parameter
$w_{\rm{eff}}$ blows up (see Eq.~(\ref{weff}) for $\alpha=0$  where
$\alpha$ is the GB parameter). ($ii$) GB effects induce  ultraviolet
corrections  on the brane \cite{GB}, as expected from high-energy
stringy features, and therefore may modify the phantom-like
behaviour at earlier times and may alleviate the shortcome mentioned
in the previous point. And ($iii$), even though the DGP model is
characterised by an interesting infrared effect of gravity occurring
with respect to general relativity, which for the self-accelerating
branch can lead to a late-time acceleration on the brane even in the
absence of any exotic matter invoked to produce the dark energy
effect \cite{Deffayet:2000uy},  it would be expected that a
consistent DGP brane model would have ultraviolet modifications as
well, associated to high-energy stringy effects at earlier times
\cite{Kofinas:2003rz}.

The phantom-like behaviour on the brane is based on ($i$) writing down the modified Friedmann equation of the brane as the standard relativistic Friedmann equation (see Eq.~(\ref{Friedmann3})) and ($ii$)
defining a corresponding effective energy density, $\rho_{\rm{eff}}$,  which
grows as the brane expands, and an effective equation of state (cf.
Eqs.~(\ref{rhoeff}) and (\ref{weff}) and Figs.~\ref{prho} and
\ref{pweff}). The effective energy density corresponds to a balance
between the cosmological constant and geometrical effects encoded on
the Hubble rate. This was done by generalising the way a
phantom-like behaviour is obtained in the $\Lambda$DGP model
\cite{Sahni:2002dx,Lue:2004za} and other of its variants
\cite{Chimento:2006ac,BouhmadiLopez:2007ts} where dark energy has a
more dynamical character. As $\rho_{\rm{eff}}$ depends explicitly on
the Hubble rate, in order to get its evolution with the redshift it
is necessary to solve the cubic Friedmann equation
(\ref{Friedmannz}). This was done in section \ref{sect4}. This
analysis has also allowed us to constraint the set of most likely
values of the model (see Fig.~\ref{p3}) and therefore to pick up the
suitable cubic solution of the Friedmann equation
(\ref{Friedmannz}).

It turns out that the phantom-like behaviour also breaks down in the
$\Lambda$DGP-GB model, namely  when $\rho_{\rm{eff}}=0$; i.e. when the
brane cosmological constant balances the geometrical effects
described in terms of the Hubble rate (see Eq.~(\ref{rhoeff})). This feature highlights that the mapping between
the $\Lambda$DGP-GB model and the 4D relativistic phantom cosmology model ceases to be valid although
the brane description remains valid. The
redshift, $z_b$, at which this event happens is given  in
Eq.~(\ref{zb}) and depends on the set
($\Omega_m$,$\Omega_\Lambda$,$\Omega_{r_c}$,$\Omega_\alpha$); i.e on
the amount of CDM, the weight of  the cosmological constant, as well
as on the weights of the curvatures effects encoded on
$\Omega_{r_c}$ and $\Omega_\alpha$  (related to the crossover scale
and the GB parameter $\alpha$, respectively).
Concerning such feature, namely that the  phantom like-behaviour in
the $\Lambda$DGP and $\Lambda$DGP-GB breaks down at the redshifts  $\tilde{z}_b$
and $z_b$, respectively, can it then be that
$z_b>\tilde{z}_b$? I.e. can the GB effect bring  the breakdown of
the phantom-like behaviour in the $\Lambda$DGP to higher redshifts?
This question has been addressed analytically (in specific
situations) as well as numerically in more general situations and the
answer has been always \emph{negative} (cf. Figs.~\ref{GB-effect}
and \ref{pzbbreak}): The GB effect seems instead to induce a
breakdown of the phantom-like behaviour in the $\Lambda$DGP scenario
at an even smaller redshift. Namely, the $\Lambda$DGP has a regular
phantom-like behaviour for $[0, \tilde{z}_b)$ whereas  the phantom-like behaviour
in the $\Lambda$DGP-GB model
 is regular only for $[0,z_b)$, with $z_b<\tilde{z}_b$.
Thus,  we conclude that the GB term does extend the regime of validity of the phantom mimicry in the
$\Lambda$DGP model.

We were expecting that new curvature corrections would modify the regime of validity  of the phantom-like behaviour on the brane. We have shown that this is the case by considering a GB term in the bulk. However, these bulk curvature corrections  rather than enlarging the regime of validity of the phantom-like behaviour on the brane, they make it smaller. It might be that we need to consider exclusively curvature corrections to the induced gravity action on the brane, modelled for example through an $f(R)$  term on the brane action \cite{frbrane}. We leave this  question for a future work.

\section*{Acknowledgements}
The authors are grateful to R. Lazkoz and Y. Shtanov for very helpful comments.
MBL is  supported by the Portuguese Agency Funda\c{c}\~{a}o para a Ci\^{e}ncia e
Tecnologia through the fellowship SFRH/BPD/26542/2006.
She also wishes to acknowledge the hospitality of the
Physics Department  of the Universidade da Beira Interior
during the completion of part of this work. This research work was supported by the grant
FEDER-POCI/P/FIS/57547/2004.

\appendix

\section{Solutions of the Friedmann equation}

The solutions of the Friedmann equation (\ref{Friedmannnb}) can be written as \cite{Abramowitz}
\begin{eqnarray}
\bar{H}_1&=&\,\,\,\,\,\,\,\,\,\,\, S_1+S_2-\frac13,\label{H1}\\
\bar{H}_2&=&-\frac12(S_1+S_2)-\frac13+i\frac{\sqrt{3}}{2}(S_1-S_2)\label{H2},\\
\bar{H}_3&=&-\frac12(S_1+S_2)-\frac13-i\frac{\sqrt{3}}{2}(S_1-S_2)\label{H3},
\end{eqnarray}
where
\begin{equation}
S_1=\left[R+\left(Q^3+R^2\right)^\frac12\right]^\frac13,\,\,\,\,
S_2=\left[R-\left(Q^3+R^2\right)^\frac12\right]^\frac13.
\end{equation}
$Q$ and $R$ are given in Eq.~(\ref{QR}). Only those solutions real and positive corresponds to cosmologically interesting solutions. The latter depends strongly on the ratio between the GB parameter $\alpha$ and the crossover scale $r_c$. Once the  real roots of Eq.~(\ref{Friedmannnb}) are identified, it is much more useful to rewrite them as trigonometric function (if all the solutions are real) or as hyperbolic functions if there is a unique real solution in order to analyse the cosmological evolution of the brane. This is performed in section III.

\section{Recovery of the $\Lambda$DGP scenario}

In this appendix, we show that only two of the solutions of the Friedmann equation of the normal DGP branch with GB effects presented in Sect. \ref{sect4} have a well defined limit when $\alpha \rightarrow 0$. These solutions correspond to Eqs.~(\ref{solrel1}) and (\ref{solrel2}). Furthermore, only the solution (\ref{solrel1}) reduces to the $\Lambda$DGP scenario when $\alpha=0$.

As it can be noticed, it is  not as straightforward to take such a  limit; i.e.  $\alpha \rightarrow 0$, of the mentioned solutions because all the dimensionless parameters involved in the problem $\bar H$, $\bar\rho$ and $b$ are proportional to $\alpha$ or $\alpha^2$. To make the task easier, we introduce the following definitions
\begin{eqnarray}
\bar\rho&=&f_1\alpha^2; \quad f_1=\frac{32}{27}\frac{\kappa_5^2}{r_c^3}\rho,\\
b&=&f_2\alpha\,; \quad \,\,f_2=\frac83\frac{1}{r_c^2},\\
\bar H&=&f_3 \alpha\,;\quad \,\, f_3=\frac83 \frac{1}{r_c}H,
\end{eqnarray}
where $f_1,f_2$ and $f_3$ do not depend on $\alpha$.

We start considering the solution (\ref{solrel1}) and we  make a series expansion of $\tan(\theta)$ at $\alpha=0$, where $\theta$ is defined in Eq.~(30),
\begin{equation}
\tan(\theta)=-\frac32\left(\sqrt{12f_1+3f_2^2}\right)\alpha + O^2(\alpha).
\end{equation}
Therefore,
\begin{equation}
\theta=\pi-\frac32\left(\sqrt{12f_1+3f_2^2}\right)\alpha + O^2(\alpha),
\end{equation}
because $0<\theta<\pi$.

On the other hand, by combining the last equation and
\begin{equation}
\cos\left(\frac{\pi}{3}-cx\right)=\frac12\left(1+\sqrt3 cx\right)  +  O^2(x)
\end{equation}
for $x\rightarrow 0$ and $c$ a constant, we obtain
\begin{equation}
\cos\left(\frac{\theta}{3}\right)=\frac12\left[1+\frac{\sqrt3}{2}\left(\sqrt{12f_1+3f_2^2}\right)\alpha\right]+
O^3(\alpha).
\end{equation}
Finally, we substitute the last equation in Eq.~(\ref{solrel1}) and we conclude
\begin{eqnarray}
f_3\alpha&=&\frac13\left\{\left(1-\frac32 f_2\alpha\right)\left[1+\frac{\sqrt3}{2}\left(\sqrt{12f_1+3f_2^2}\right)\alpha\right]-1\right\}\nonumber \\&+&\ldots,\nonumber \\
&=&\frac12f_2\alpha\left(-1+\sqrt{1+4\frac{f_1}{f_2^2}}\right) +O^2(\alpha).
\end{eqnarray}
We equate the lower order in $\alpha$ of the last equation (of both handsides of the equation) and  we substitute the definitions of $f_1,f_2$ and $f_3$. We finally obtain
\begin{equation}
H=\frac{1}{2r_c}\left(-1+\sqrt{1+\frac23\kappa_5^2r_c\rho}\right),
\end{equation}
which is the modified Friedmann equation of the non-self-accelerating DGP branch that can be obtained from Eqs (1) or (3)
(with - sign)  with $\alpha=0$ and contains the $\Lambda$DGP model as a specific solution.

Following a similar procedure, it can be shown that the limit $\alpha=0$ of Eq.~(\ref{solrel2})
\begin{equation}
H=-\frac{1}{2r_c}\left(1+\sqrt{1+\frac23\kappa_5^2r_c\rho}\right).
\end{equation}
The limit is well defined although the solution it is not physical (at least it is not suitable for the late-time evolution of the universe). Please notice that Eq.~(3) of the manuscript  with $\alpha=0$ is quadratic on the Hubble rates and therefore has two roots. This second root; i.e. Eq. (B10), after a time reversal ($t$ by $-t$)  corresponds to the self-accelerating DGP solution.

On the other hand, the solution (32) of the paper has not a well defined $\alpha=0$ limit because the left hand side of the equation is proportional to $\alpha$ while the right handside lower order in $\alpha$ is a constant different from zero.

If one tries to take the limit $\alpha\rightarrow 0$ of the solution presented in Eq.~(25) it turns out that $\eta$ must be complex. In fact, in that case one is getting an analytical extension of the solution which corresponds to the solution (\ref{solrel1}). The reason behind this behaviour is that the solutions (\ref{solrel1}) and (25) are related by a sort of ``Wick rotation'' where $\theta=i\eta$. Notice at this respect that the function $\mathcal{N}$ changes sign and the prolongation is well defined.

For the rest of solutions presented in section III, although all of them comes from the general solution (A1), the limit $\alpha\rightarrow 0$ cannot be taken because $1/4<b$.

In summary, the recovery of the  $\Lambda$DGP scenario from the normal DGP branch with GB effects requires that ($i$) the value of the parameter $b$, which is proportional to $\alpha$, to be such that $0<b<1/4$ and ($ii$) the dimensionless amount of the brane total energy density has to fullfill $\bar\rho<\bar\rho_1$ . Therefore, in order to recover the $\Lambda$DGP scenario from the model we propose, the GB parameter has to be  bounded from above and the brane total energy density has also to be  bounded by a maximum threshold. Along section IV, we have restricted to this solution as it is the most favourable solution observationally because the parameter $b$ is expect to be small which implies that the most likely set of cosmological parameters that characterises the model would lie on the bottom of the left handside corner of figure 2. This last feature on itself implies that the brane energy density cannot be too large, more precisely $\bar\rho_{\rm{min}}\leq\bar\rho_1$ which implies in particular  that there is a range of values of $\bar\rho$ such that $\bar\rho_1$.

\end{document}